\documentclass[final,1p,times]{elsarticle}

 \usepackage{graphics}
 \usepackage{array}

\usepackage{amsmath}
 \usepackage{amsthm}

 \usepackage{lineno}

 \biboptions{comma,square,compress}

\begin{document}
\begin{frontmatter}



\title{Temperature of the hot $\alpha$-source: a guidance to probe the Bose-Einstein condensation of $\alpha$ clusters in the heavy-ion collision}

\author[1]{Jun Su \corref{cor2}}
\author[1]{Long Zhu}
\address[1]{Sino-French Institute of Nuclear Engineering and Technology, Sun Yat-sen University, Zhuhai 519082, China}

\cortext[cor2]{Corresponding author. sujun3@mail.sysu.edu.cn}

\begin{abstract}
Recent experimental efforts were made to explore the Bose-Einstein condensation in multi-$\alpha$ system using the heavy-ion collisions.
This Letter provides a explanation why no signatures were observed and a guidance for future experiments.
More precisely, a harmonic oscillator model is developed to study the multi-$\alpha$ source at or near zero temperature.
The temperature and particle population of the multi-$\alpha$ sources observed in the current experiments are extracted.
It is found that almost no $\alpha$ particles occupy the ground state in those multi-$\alpha$ sources.
The critical temperature for the multi-$\alpha$-condensed states are
predicted, which provides a guidance for the future experiments to probe the Bose-Einstein condensation of $\alpha$ clusters in the heavy-ion collisions.
\end{abstract}

\begin{keyword}

Bose-Einstein condensation, $\alpha$ cluster, Heavy-ion collision
\end{keyword}

\end{frontmatter}




The Bose-Einstein condensation (BEC) in cold many-body quantum system that consists of bosons was predicted in 1924 and 1925 by S. N. Bose and A. Einstein respectively \cite{Bose1924,Einstein1925}.
It was experimentally confirmed seventy years later in dilute atomic gases by using developments in cooling and trapping techniques \cite{anderson1995observation,davis1995bose}.
Currently, the BEC is the subject of intense in many physics fields, such as quark-gluon plasma physics \cite{xu2015thermalization}, molecular physics \cite{zwierlein2003observation}, astrophysics \cite{levkov2018gravitational}, as well as nuclear physics.

As far as nuclear physics is concerned, nuclei are normally treated as consisting of fermions.
However, due to the strong four-particle (two protons and two neutrons) correlations, the $\alpha$ cluster behaves as a well-established subunit and exhibits bosonic properties in the nuclear system at sufficiently low density \cite{carstoiu2009saturation}.
This is the basis of the theoretical prediction of the nuclear BEC, which has been put forward over twenty years ago \cite{ropke1998four,beyer2000alpha}.
It has been shown that the BEC phase plays a decisive role in understanding the phase diagram of the nuclear matter and studying the nuclear equation of state \cite{satarov2019phase,mishustin2019condensation,zhang2019cold,ebran2020alpha,satarov2020possible}.
For the experimental confirmation, its finite-size counterpart, i.e. the $\alpha$-cluster condensed state (ACS) in the light self-conjugate nuclei attracts much attention.

The $\alpha$-cluster formation in the light self-conjugate nuclei was predicted in the 1930s \cite{bethe1936nuclear,wefelmeier1937geometrisches}, and then experimentally observed at excitation energies in the vicinity of the $\alpha$ emission threshold \cite{freer2018microscopic}.
Theoretical calculations supported that the $\alpha$ clusters in the ground state of $^{8}$Be \cite{mohr1994properties,funaki2002description}, 0$^{+}_{2}$ state of $^{12}$C \cite{tohsaki2001alpha,funaki2003analysis,uegaki1979structure,kamimura1981transition,zhao2015rod}, and 0$^{+}_{6}$ state of $^{16}$O \cite{funaki2008alpha,yamada2012isoscalar,funaki2018container} condense into the same lowest 0s orbit, which provide indirect evidences of 2$\alpha$, 3$\alpha$, and 4$\alpha$ condensed states, respectively.
Candidates for the 5$\alpha$ condensed state in $^{20}$Ne has also been suggested \cite{adachi2021candidates}.
For heavier conjugate nuclei, the molecular structures have been investigated theoretically \cite{wuosmaa1992evidence,anagnostatos1998alpha,ito2002multicluster,yamada2004dilute,wang2011alpha,royer2015energies, inakura2018rod}, but no 0$^{+}$ states are assigned to the $\alpha$-particle condensation.

The experimental signal of the BEC in nuclear system is of relevance in heavy-ion collisions at a relatively late stage where the temperature of expanding matter has dropped \cite{ropke1998four}.
To this end, the multi-$\alpha$ sources were verified in the heavy-ion collisions \cite{zhang2019strongly,huang2021four}.
The excitation energy distributions were measured for 5$\alpha$ produced in $^{16}$O + $^{12}$C at 10, 17.5, and 25 MeV/nucleon \cite{bishop2019experimental}, 6$\alpha$ produced in central $^{12}$C + $^{12}$C reactions at 7.9 MeV/nucleon \cite{morelli2019full}, 7$\alpha$ produced in the collisions $^{28}$Si + $^{12}$C at 35 MeV/nucleon \cite{cao2019examination}, and 10$\alpha$ source in collisions $^{40}$Ca + $^{40}$Ca at 35 MeV/nucleon \cite{schmidt2017alpha}.
The fragmentation of quasi-projectiles from the nuclear reaction $^{40}$Ca + $^{12}$C at 25 MeV/nucleon was also used to produce 4$\alpha$, 5$\alpha$, and 6$\alpha$ sources, from which the particle spectra were measured \cite{borderie2016probing,raduta2011evidence}.
To provide the direct signatures for the $\alpha$ condensation, a key question is how many $\alpha$ clusters in those hot $\alpha$-sources occupy the lowest orbit.
The aim of the present Letter is to extract the temperatures of the multi-$\alpha$ sources observed in the current experiments and explore how close we were to BEC of the $\alpha$ clusters.

A particle interacting with other same particles can be described approximatively by solving the Schrodinger's equation \cite{goldhammer1963structure},
  \begin{equation}
    \left [ -\frac{\hbar^{2}}{2M}\nabla^{2} +U \right]\psi = E\psi,
  \label{Sch_Equ}
  \end{equation}
where $U$ represents the interaction between the particle and the residues, and $M$ is the reduced mass,
  \begin{equation}
    M = \frac{k-1}{k}m_{\alpha},
  \label{mu}
  \end{equation}
$m_{\alpha}$ is the rest mass of the particle, and $k$ is the particle number in the system.
In the three-dimensional harmonic oscillator potential with parameters $\omega$ and $V$,
  \begin{equation}
    U = \frac{1}{2}M^{2}\omega^{2}-V,
  \label{hop}
  \end{equation}
there are analytical solutions.
The energy eigenvalues are,
  \begin{equation}
  \begin{aligned}
    E_{i} &= (i+\frac{3}{2})\hbar \omega -V, \\
    i &= 2n+l,
  \label{En}
  \end{aligned}
  \end{equation}
where $i$ is the number of the energy level, $n$ is the principle quantum number, $l$ is the orbital angular momentum number.
Sine there is other quantum number, i.e. the magnetic number m which is not expressed in Eq. (\ref{En}), the degeneracy of energy levels is,
  \begin{equation}
    g_{i} = \frac{1}{2}(i+1)(i+2).
  \label{deg}
  \end{equation}

For the Bose system at zero temperature, all particles occupy in the first level.
It is the ground state, or called the Bose-Einstein condensation state, where the wave function and energy of the particle are,
  \begin{equation}
  \begin{aligned}
    \psi(r, \theta, \phi) &= \left(\frac{M\omega}{\pi\hbar}\right)^{3/4} \exp\left(-\frac{M\omega }{2\hbar}r^{2}\right), \\
    E_{0} &= \frac{3}{2}\hbar \omega -V.
  \label{wf}
  \end{aligned}
  \end{equation}
Then, the particle density and energy of the system can be calculated, 
  \begin{equation}
  \begin{aligned}
     \rho(r) &= k \left(\frac{M\omega}{\pi\hbar}\right)^{3/2} \exp\left(-\frac{M\omega }{\hbar}r^{2}\right),\\
     E_{s0} &= k (\frac{3}{2}\hbar\omega -V).
  \label{rms}
  \end{aligned}
  \end{equation}

For the case at finite temperature T, the particle population $n_{i}$($E_{i}$) of $i$-th state with energy $E_{i}$ is given by the Bose-Einstein distribution,
  \begin{equation}
     n_{i}(E_{i}) = \frac{g_{i}}{\exp(\frac{E_{i}-\mu}{T})-1},
  \label{n_Ei}
  \end{equation}
where $\mu$ is the chemical potential.
It is temperature dependence and can be determined by the constraint of the particle number,
  \begin{equation}
     \sum_{i=0}^{\infty} n_{i} = k.
  \label{mu}
  \end{equation}
The energy of the system is calculated by,
  \begin{equation}
     E_{s}(T) = \sum_{i=0}^{\infty} n_{i} E_{i}.
  \label{E_T}
  \end{equation}

This harmonic oscillator model can be applied to study approximatively the multi-$\alpha$ system at or near zero temperature.
It is assumed that the center density $\rho_{\alpha0}$ of the conjugate nuclei in the ACS is independent of the particle number $k$, then one can replace the parameter $\omega$ by a new one $\omega_{0}$, and $V$ by $V_{0}$,
  \begin{equation}
  \begin{aligned}
    \omega &= \frac{k^{1/3}}{k-1} \omega_{0}, \\
    V &= \frac{V_{0}}{k-1}.
  \label{omiga0}
  \end{aligned}
  \end{equation}
The problem may now be determination of the parameters $\omega_{0}$ and $V_{0}$.

\begin{figure}[htb]
	\centering
	\includegraphics[width=8.5cm]{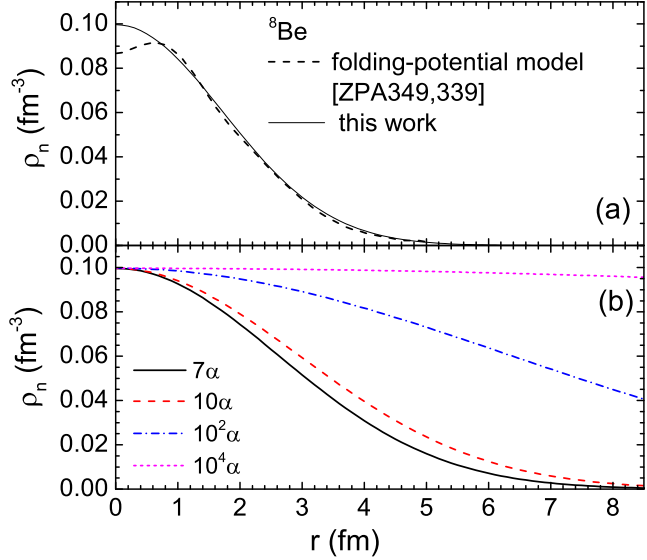}
	\caption{Nucleon density distributions in multi-$\alpha$ system at zero temperature calculated by the harmonic oscillator model. In panel(a), the case in the 2$\alpha$ system ($^{8}$Be) is compared with the calculation by the folding-potential model, which is taken from Ref. \cite{mohr1994properties}. Panel(b) displays the cases for 7$\alpha$, 10$\alpha$, 10$^{2}$$\alpha$, and 10$^{4}$$\alpha$ systems. The 10$^{4}$$\alpha$ system is considered approximatively the infinite $\alpha$ matter with nucleon density 0.1 fm$^{-3}$ ($\alpha$ density 0.025 fm$^{-3}$). }
	\label{fig-1}
\end{figure}

Two-$\alpha$ cluster structure is well known in the $^{8}$Be ground state, which has been calculated by several models.
For example in the folding-potential model \cite{mohr1994properties}, the $\alpha$-$\alpha$ potential is obtain by fitting the experimental differential cross section in $\alpha$-$\alpha$ scattering as well as the resonance energy and width of the $^{8}$Be ground-state.
The nucleon density distribution is then calculated by the s-wave function for the relative motion of two $\alpha$-clusters.
Setting the width parameter in the harmonic oscillator potential as $\hbar\omega_{0}$ = 2.78 MeV, the nucleon density distribution calculated in this work agrees to that by the folding-potential model, shown as in Fig. \ref{fig-1}(a).
A slight difference can be seen at the center.
The center nucleon density 0.1 fm$^{-3}$ of the conjugate nuclei keeps constant but the distribution tends to widen with increasing particle number $k$, see Fig. \ref{fig-1}(b).
Setting $k$ $\rightarrow$ $\infty$, one obtains the infinite $\alpha$ matter with $\alpha$ density 0.025 fm$^{-3}$ (a quarter of the nucleon density).
It is in the proposed range of the realistic saturation point of the $\alpha$ matter, i.e. 0.015-0.0325 fm$^{-3}$ \cite{carstoiu2009saturation}.

\begin{figure}[htb]
	\centering
	\includegraphics[width=8.5cm]{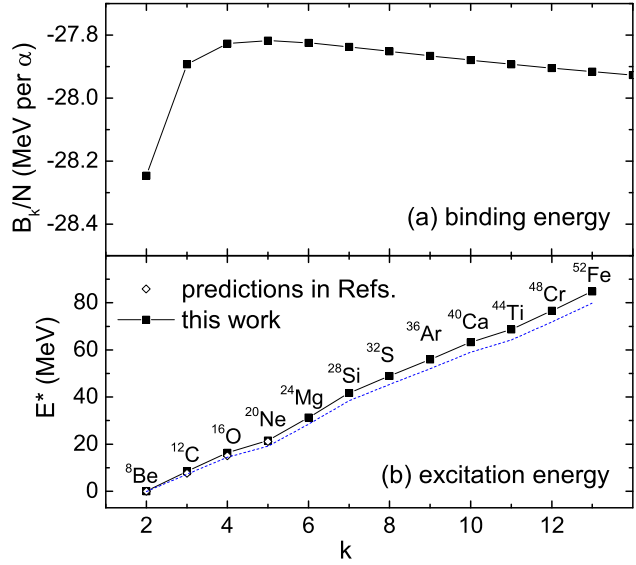}
	\caption{(a) Binding energies of the conjugate nuclei in the ACS. (b) Excitation energies of the conjugate nuclei in the ACS calculated in this work and predicted in Refs. \cite{uegaki1979structure, kamimura1981transition, Funaki2003} for $^{12}$C , \cite{funaki2008alpha, yamada2012isoscalar, funaki2018container} $^{16}$O, and \cite{yamada2004dilute} for $^{20}$Ne. The threshold energies of the ACSs are shown as dashes.}
	\label{fig-2}
\end{figure}

The binding energy and excitation energy of the conjugate nucleus in the ACS is calculated by
  \begin{equation}
  \begin{aligned}
    B_{k} &= E_{s0} +kB_{\alpha}, \\
    E^{*} &= B_{k} -B,
  \label{E*}
  \end{aligned}
  \end{equation}
where $B_{\alpha}$ = -28.30 MeV is the binding energy of $\alpha$ particle, and $B$ is the binding energy of the conjugate nucleus in the ground state.
Using the binding energy of $^{8}$Be ($B_{2}$ = -56.50 MeV), we determine the depth parameter of the potential, i.e. $V_{0}$ = 5.20 MeV.
Then the cases for other conjugate nucleus can be calculated, as shown in Fig. \ref{fig-2}.
With increasing $\alpha$-particle number, the binding energy per $\alpha$-particle raises and falls with the maximum at $k$ = 5.
This shape is similar to that of the binding energy of nucleus, and associated with the saturability of the nuclear force.

The excitation energies of the conjugate nuclei in the ACSs are shown as a function of $\alpha$-particle number $k$ in Fig. \ref{fig-2}(b).
An approximately linear increasing trend can be seen.
Comparing with the threshold energies of the ACSs (shown as dashes), the excitation energies are slightly larger.
This is consistent with the fact that the ACSs in the conjugate nuclei are probed near and above the thresholds.
The $0^{+}$ states at 7.65 MeV for $^{12}$C \cite{morinaga1956interpretation}, 15.10 MeV for $^{16}$O \cite{tilley1993energy}, and 21.2 MeV for $^{20}$Ne \cite{adachi2021candidates} have been measured experimentally.
The strong evidence of the ACS in $^{12}$C has been provided by the alpha-cluster model \cite{uegaki1979structure,kamimura1981transition,funaki2003analysis} and density functional theory \cite{zhao2015rod}.
Theoretical calculations for $^{16}$O \cite{funaki2008alpha,yamada2012isoscalar,funaki2018container} and $^{20}$Ne \cite{yamada2004dilute,adachi2021candidates} were also reported.
Our predictions for those three nuclei are $E^{*}$ = 8.48, 16.3, and 21.6 MeV, which are larger slightly than those in the references.

\begin{figure}[htb]
	\centering
	\includegraphics[width=8.5cm]{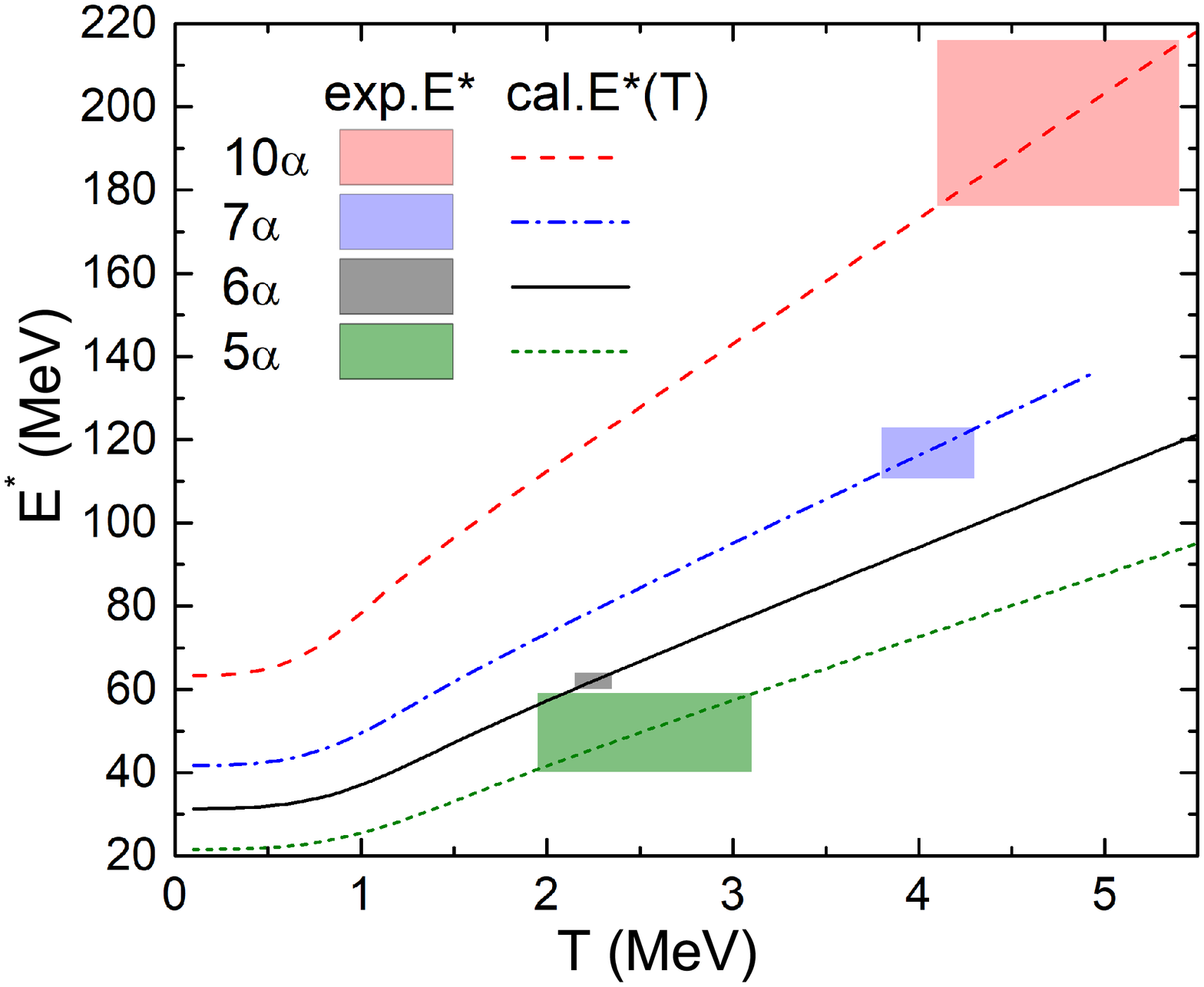}
	\caption{Excitation energies of the $k\alpha$ states as a function of the temperature, for $k$ = 5, 6, 7, and 10. The $k\alpha$ states at excitation energy 49$\pm$8 MeV for 5$\alpha$ \cite{bishop2019experimental}, 62$\pm$2 MeV for 6$\alpha$ \cite{morelli2019full}, 117$\pm$6 MeV for 7$\alpha$ \cite{cao2019examination}, and 196$\pm$20 MeV for 10$\alpha$ \cite{schmidt2017alpha} have been measured in experiments, shown as shadow. Comparing the calculated $E^{*}$-T curves with the measured excitation energies, it is extracted the temperature of the measured $k\alpha$ states, 2.47$\pm$0.47 MeV for 5$\alpha$, 2.25$\pm$0.10 MeV for 6$\alpha$, 4.05$\pm$0.25 MeV for 7$\alpha$, and 4.75$\pm$0.65 MeV for 10$\alpha$.}
	\label{fig-3}
\end{figure}

\begin{figure}[htb]
	\centering
	\includegraphics[width=8.5cm]{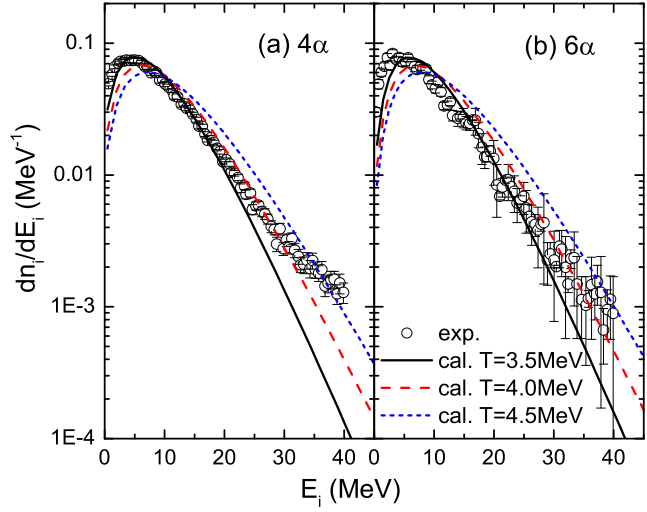}
	\caption{Particle spectra from $4\alpha$ and $6\alpha$ states. Curves display the calculations within the temperature T = 3.5, 4.0, and 4.5 MeV. The circles show the experimental data taken from  \cite{borderie2016probing}.}
	\label{fig-4}
\end{figure}

\begin{table}[tbp]
   \centering
   \caption{\label{T}
   Extracted temperatures of the multi-$\alpha$ sources produced in heavy-ion collisions by the  harmonic oscillator (HO) model and isospin-dependent quantum molecular dynamics (IQMD) model. The extraction method by the IQMD model is demonstrated in our previous work \cite{su2018fusion}. }
   \begin{tabular}{p{1.0cm}<{\raggedright} p{1.2cm}<{\centering} p{1.3cm}<{\centering} p{1.5cm}<{\centering} p{0.6cm}<{\centering}}
   \hline
   \hline
     \multicolumn{2}{c}{Heavy-ion collision}  & & \multicolumn{2}{c}{T(MeV)}  \\
    system & E(MeV/n) &$\alpha$-source  & HO model & IQMD \\
   \hline
   $^{16}$O+$^{12}$C & 10 & 5$\alpha$ & 2.47$\pm$0.47 &   \\
   $^{12}$C+$^{12}$C & 7.9 & 6$\alpha$ & 2.25$\pm$0.10 &   \\
   $^{28}$Si+$^{12}$C & 35 & 7$\alpha$ & 4.05$\pm$0.25 & 5.0 \\
   $^{40}$Ca+$^{40}$Ca & 35 & 10$\alpha$ & 4.75$\pm$0.65 & 5.5 \\
   $^{40}$Ca+$^{12}$C & 25 & 4$\alpha$/6$\alpha$ &about 4 & 4.5 \\
   \hline
   \hline
\end{tabular}
\end{table}

Using the excitation energy distributions taken from Refs. \cite{bishop2019experimental, morelli2019full, cao2019examination, schmidt2017alpha}, it is calculated the average and variance as 49$\pm$8 MeV for 5$\alpha$, 62$\pm$2 MeV for 6$\alpha$, 117$\pm$6 MeV for 7$\alpha$, 196$\pm$20 MeV for 10$\alpha$.
Particle spectra from multi-$\alpha$ sources are also measured using $^{40}$Ca + $^{12}$C collisions at 25 MeV/nucleon \cite{borderie2016probing}. 
We calculate the excitation energy as a function of the temperature (Fig. \ref{fig-3}) and the particle spectra at different temperatures (Fig. \ref{fig-4}).
Comparing the calculations with the experimental data, it can be extracted the temperature of the multi-$\alpha$ sources produced in the heavy-ion collisions, as shown in Table \ref{T}.
The temperature of the colliding system at the moment when it reaches the nucleon density $\rho_{n}$ = 0.1 fm$^{-3}$ are extracted by the isospin-dependent quantum molecular dynamics (IQMD) model \cite{su2018fusion}, as shown in the last column in the Table.
The calculations by the IQMD model are larger than those by the harmonic oscillator model with 1 MeV.

\begin{figure}[htb]
	\centering
	\includegraphics[width=8.5cm]{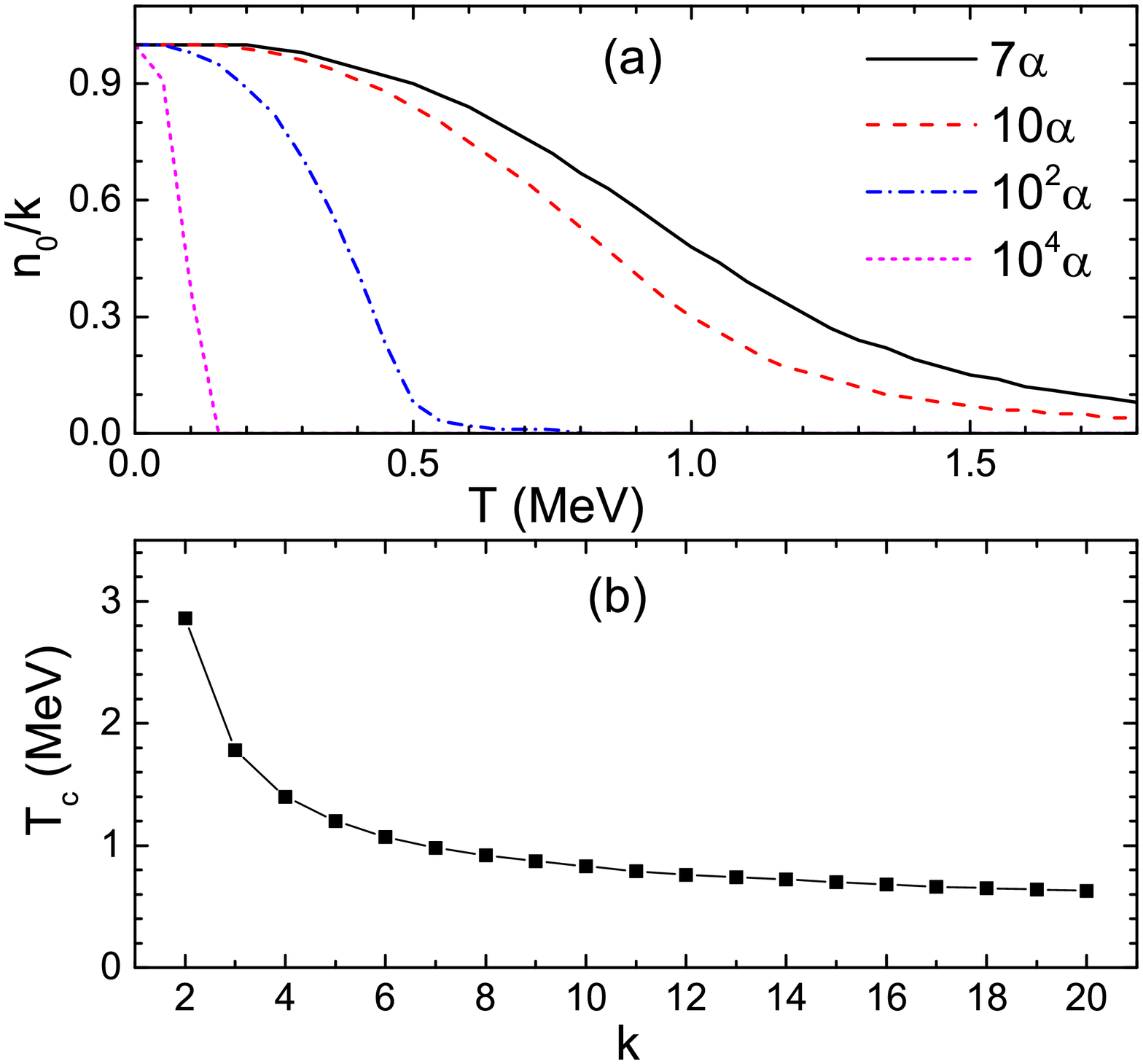}
	\caption{(a) Condensate fraction of the $\alpha$ particle as a function of the temperature in the 7$\alpha$, 10$\alpha$, 10$^{2}$$\alpha$ and 10$^{4}$$\alpha$ systems. (b) Critical temperatures of the multi-$\alpha$ system for its condensate fraction is 0.5.}
	\label{fig-5}
\end{figure}

Do we assign these measured multi-$\alpha$ events to the ACSs?
To answer this question we plot in Fig. \ref{fig-5}(a) the condensate fraction of the $\alpha$ particle as a function of the temperature in the 7$\alpha$, 10$\alpha$, 10$^{2}$$\alpha$ and 10$^{4}$$\alpha$ systems.
The condensate fraction is defined as the particle population at the ground state reduced to the $\alpha$-particle number $k$ of the systems.
For the infinite system, the condensate fraction is 1 only when the temperature is approaching to zero.
The 10$^{4}$$\alpha$ system is similar to this case.
However for the finite system, one can observe a considerable condensate fraction at finite temperature \cite{kakanis1996persisting}.
Taking the 7$\alpha$ system for example, the condensate fraction is 0.9 at T = 0.5 MeV, and decreases to 0.48 at T = 1.0 MeV.
The condensate fraction is falling faster for the heavier system.
For the 10$\alpha$ system, almost no $\alpha$ particles occupy the ground state when the temperature is larger than 1.5 MeV, which is much smaller than the extracted temperatures in Table \ref{T}.

We definite the critical temperature $T_{c}$ of the multi-$\alpha$ system at which 50\% of the $\alpha$ particles occupy the ground state, i.e. the condensate fraction is 0.5.
It is proposed that the colliding nuclear system should be cooled to less than $T_{c}$, or else the $\alpha$ particle condensation can not be observed.
The condensate temperature $T_{c}$ as a function of the $\alpha$ number $k$ is shown in Fig. \ref{fig-5}(b).
The critical temperature of the 5$\alpha$, 6$\alpha$, 7$\alpha$, and 10$\alpha$ systems are 1.20, 1.07, 0.98, and 0.83 MeV, which are much smaller than the extracted temperatures of the multi-$\alpha$ sources observed in recent experiments (see Table \ref{T}).
This is the reason why no experimental signatures for $\alpha$ condensation were observed in those experiments.
Lower temperature should be reached when probing the $\alpha$ particle condensation in the heavy-ion collision.
Lower temperature means less incident energy, see Fig. 2 in Ref. \cite{su2018fusion}.
However, one should pay attention to the fact that the nuclear system at the condensate density (0.1 fm$^{-3}$) can not be produced by the heavy-ion collisions at too small incident energy.


In summary, a harmonic oscillator model is applied to study the multi-$\alpha$ system at or near zero temperature.
This model only includes two parameters but successfully reproduces the nucleon density distribution in $^{8}$Be, excitation energies of the conjugate nuclei $^{8}$Be, $^{12}$C, $^{16}$O, and $^{20}$Ne, particle spectra from 4$\alpha$ and 6$\alpha$ sources, and the realistic saturation density of the infinite $\alpha$ matter.
Using the excitation energies and particle spectra observed in the current experiments \cite{bishop2019experimental, morelli2019full, cao2019examination, schmidt2017alpha, borderie2016probing}, the temperatures of the multi-$\alpha$ sources produced in the heavy-ion collisions are extracted.
It is shown that almost no $\alpha$ particles occupy the lowest orbit at those temperatures.
This is the reason why no experimental signatures for $\alpha$ condensation were observed in those experiments.
The critical temperature of the multi-$\alpha$-condensed states are predicted.
This work provides a guidance for the future experiments to probe the Bose-Einstein condensation of $\alpha$ clusters in the heavy-ion collisions.

\section*{ACKNOWLEDGMENTS}

This work was supported by the National Natural Science Foundation of China under Grants Nos. 11875328 and 12075327.


\bibliographystyle{elsarticle-num-names}
\bibliography{bibfile_alpha}

\end{document}